\documentclass[12pt]{article}
\usepackage{epsfig}
\usepackage{subfigure}
\usepackage{amsmath}
\usepackage{amsfonts}
\usepackage{amssymb}
\usepackage{url}
\usepackage{hyperref}
\newcommand{\be}{\begin{equation}}
\newcommand{\ee}{\end{equation}}
\newcommand{\bea}{\begin{eqnarray}}
\newcommand{\eea}{\end{eqnarray}}

\begin{document}
\begin{titlepage}
\begin{flushright}
\end{flushright}
\vspace{4\baselineskip}
\begin{center}
{\Large\bf 
A simple SO(10) GUT in five dimensions 
}
\end{center}
\vspace{1cm}
\begin{center}
{\large Takeshi Fukuyama $^{a,}$
\footnote{\tt E-mail:fukuyama@se.ritsumei.ac.jp} 
and Nobuchika Okada $^{b,}$
\footnote{\tt E-mail:okadan@post.kek.jp}}
\end{center}
\vspace{0.2cm}
\begin{center}
${}^{a}$ {\small \it Department of Physics, Ritsumeikan University,
Kusatsu, Shiga, 525-8577, Japan}\\[.2cm]
${}^{b}$ {\small \it Theory Division, KEK,
Oho 1-1, Tsukuba, Ibaraki, 305-0801, Japan}\\
\medskip
\vskip 10mm
\end{center}
\vskip 10mm
\begin{abstract}
A simple supersymmetric SO(10) GUT in five dimensions is considered. 
The fifth dimension is compactified on the $S^1/(Z_2\times Z_2^\prime)$ 
 orbifold possessing two inequivalent fixed points. 
In our setup, all matter and Higgs multiplets reside 
 on one brane (PS brane) where the original SO(10) gauge group 
 is broken down to the Pati-Salam (PS) gauge group, 
 SU(4)$_c \times$SU(2)$_L \times$ SU(2)$_R$, 
 by the orbifold boundary condition, 
 while only the SO(10) gauge multiplet resides in the bulk. 
The further breaking of the PS symmetry to the Standard Model 
 gauge group is realized by Higgs multiplets on the PS brane 
 as usual in four dimensional models. 
Proton decay is fully suppressed. 
In our simple setup, the gauge coupling unification 
 is realized after incorporating 
 threshold corrections of Kaluza-Klein modes. 
When supersymmetry is assumed to be broken on the other brane, 
 supersymmetry breaking is transmitted to 
 the PS brane through the gaugino mediation  
 with the bulk gauge multiplet.

\end{abstract}
\end{titlepage}

\setcounter{page}{1}
\setcounter{footnote}{0}

\section{Introduction}

Current experimental data for the Standard Model (SM) 
 gauge coupling constants suggest 
 the successful gauge coupling unification 
 in the minimal supersymmetric (SUSY) Standard Model (MSSM) 
 and thus strongly support the emergence of a SUSY GUT 
 around $M_{\rm GUT} \simeq 2 \times 10^{16}$ GeV.

Among several GUTs, a model based on the gauge group SO(10) 
 is particularly attractive. 
In fact, SO(10) is the smallest simple gauge group 
 under which the entire SM matter content of each generation 
 is unified into a single anomaly-free irreducible representation, 
 ${\bf 16}$ representation. 
This ${\bf 16}$ representation automatically includes 
 the right-handed neutrino and the SO(10) GUT incorporates 
 the see-saw mechanism \cite{see-saw} 
 that can naturally explain the lightness of the light neutrino masses.

Among several models based on the gauge group SO(10), 
 the so-called renormalizable minimal SO(10) model has been paid 
 a particular attention, where two kinds of Higgs multiplets 
 $\{{\bf 10} \oplus {\bf \overline{126}}\}$ 
 are utilized for the Yukawa couplings with matters 
 ${\bf 16}_i~(i=\mbox{generation})$ 
 \cite{Babu:1992ia, Bajc:2002iw}%
\footnote{
On the other hand, there is another flow of 
 non-renormalizable minimal SO(10) GUT \cite{Chang:2004pb}. 
}. 
A remarkable feature of the model is its high predictivity 
 for the neutrino oscillation data 
 as well as charged fermion masses and mixing angles. 
After KamLAND data \cite{Eguchi:2002dm} was released, 
 it entered to the precise calculation phase, 
 and many authors performed new data fitting 
 to match up these new data \cite{Goh:2003hf}.

High predictivity of renormalizable SUSY SO(10) model was shown 
 in constructing a concrete Higgs sector of the minimal SO(10) model. 
A simplest and renormalizable Higgs superpotential 
 was constructed explicitly and the patterns 
 of the SO(10) gauge symmetry breaking to the Standard Model one 
 was shown \cite{Fukuyama:2004xs, Bajc:2004xe}.
This construction gives some constraints among the vacuum expectation
 values (VEVs) of several Higgs multiplets, 
 which give rise to a trouble in the gauge coupling unification. 
The trouble comes from the fact that the observed neutrino oscillation 
 data suggests the right-handed neutrino mass around 
 $10^{12-14}$ GeV, which is far below the GUT scale. 
This intermediate scale is provided by Higgs field VEV, 
 and several Higgs multiplets are expected to have their masses 
 around the intermediate scale and contribute to 
 the running of the gauge couplings. 
Therefore, the gauge coupling unification 
 at the GUT scale may be spoiled. 
This fact has been explicitly shown in Ref.~\cite{Bertolini:2006pe}, 
 where the gauge couplings are not unified any more 
 and even the ${\rm SU}(2)$ gauge coupling blows up below the GUT scale. 
In order to avoid this trouble and keep the successful gauge coupling 
 unification as usual, we have several choices.
One conservative approach is to add ${\bf 120}$ Higgs 
 and we may adjust newly introduced parameters 
 so as to unify gauge couplings.

In addition to the issue of the gauge coupling unification, 
 the minimal SO(10) model potentially suffers from the problem 
 that the gauge coupling blows up around the GUT scale. 
This is because the model includes many Higgs multiplets of 
 higher dimensional representations. 
In field theoretical point of view, this fact implies 
 that the GUT scale is a cutoff scale of the model, 
 and more fundamental description of the minimal SO(10) model 
 would exist above the GUT scale. 
As a simple realization of such a scenario, 
 we have considered the minimal SO(10) model 
 in a warped extra dimension \cite{F-K-O}. 
In this scenario, the AdS curvature and the fifth dimensional 
 radius were chosen so as to realize the GUT scale 
 as an effective cutoff scale in 4D effective theory 
 via the warped metric \cite{RS}. 
Furthermore, we have shown that in this context, 
 the right-handed neutrino mass scale can be geometrically 
 suppressed by a few order of magnitudes from the GUT scale, 
 leaving Higgs field VEVs at the GUT scale. 
Thus, the gauge coupling unification remains as usual in the MSSM. 
This idea has been utilized in an extended model proposed 
 in Ref.~\cite{MOY}, where the so-called type II seesaw mechanism 
 dominates to realize the tiny neutrino masses 
 through the warped geometry.

In these extra-dimensional SO(10) models, 
 it is assumed that the SO(10) gauge symmetry breaking 
 is correctly achieved by Higgs multiplets on a brane 
 as usual in 4D model. 
In addition, to realize non-trivial wave functions of matters 
 in the bulk, non-zero VEV of the chiral adjoint multiplet 
 ${\bf 45}$ in the bulk N=2 SUSY gauge multiplet 
 is assumed, which breaks SO(10) into SU(5)$\times$U(1)$_X$. 
Since the $Z_2$ orbifold parity for this adjoint multiplet 
 is assigned as odd, its VEV induces Fayet-Iliopoulos D-terms 
 localized on the branes at the orbifold fixed points 
 \cite{D-term}, which should be canceled out 
 by some Higgs multiplets on the branes, 
 in order to preserve SUSY. 
In this point, it may be not impossible but more complicated 
 to construct a model including a complete Higgs sector 
 in this class of extra-dimensional GUT models.

In this paper we consider another possibility 
 for constructing extra-dimensional GUT models, 
 the orbifold GUT \cite{Kawamura, Altarelli, Koba, Hall, Hebecker}. 
In this context, the GUT gauge symmetry is broken 
 by the orbifold boundary condition without Higgs multiplets. 
This boundary conditions can also realize the triplet-doublet 
 Higgs boson mass splitting by projecting out the zero-mode 
 of triplet Higgs while leaving the doublet Higgs one. 
In addition, dangerous dimension five operators 
 causing rapid proton decay can also be projected out, 
 the mechanism of which can be interpreted
 in terms of R-symmetry in 4D theoretical point of view. 
There are so many papers in the context of the orbifold GUT.

We propose, in this paper, a simple and clearcut scenario 
 based on a SUSY SO(10) model in five dimensions. 
Usually, orbifold SO(10) models were considered in six dimensions \cite{6D}, 
 because we need at least two projections for SO(10) down 
 to SM gauge group \cite{JEKim} 
 if we break the symmetry only through boundary conditions. 
In our scenario, the fifth dimension is compactified on 
 the $S^1/(Z_2\times Z_2^\prime)$ orbifold \cite{Kawamura, Altarelli, Hall}, 
 which has two inequivalent fixed points. 
By the orbifold boundary conditions, 
 a bulk N=2 SUSY (in the sense of 4D) and SO(10) gauge symmetry  
 are broken down to N=1 SUSY PS model with 
 the gauge group SU(4)$_c\times$SU(2)$_L\times$SU(2)$_R$. 
Further gauge symmetry breaking to the SM gauge group 
 is achieved in the usual 4D manner by VEVs of suitable Higgs 
 multiplets on a brane. 
This class of SO(10) models have been proposed 
 by several authors \cite{Raby} and some improvements 
 compared to the 6D models have been pointed out. 
Except for a common feature that the SO(10) gauge multiplet 
 resides in the 5D bulk, there are many possibilities on which 
 matter and Higgs multiplets are placed in the bulk or on one 
 of the branes at the orbifold fixed points \cite{Raby}. 
In our model, all matter and Higgs multiplets reside only 
 on one brane where the PS gauge symmetry is manifest (PS brane), 
 and thus the basic structure on the PS brane is the same 
 as the 4D PS model. 
The PS gauge symmetry is broken to the SM one by VEVs of 
 some Higgs multiplets on the brane. 
As in 4D PS models, there is no proton decay problem 
 induced by the dimension 5 operators \cite{Mohapatra}. 
With simple particle contents, we show that the gauge 
 coupling unification is realized at $M_{\rm GUT} = 4.6 \times 10^{17}$ GeV, 
 where a more fundamental theory is assumed to take place, 
 with the compactification scale at $M_c=1.2 \times 10^{16}$ GeV. 
When we assume SUSY breaking on the other brane, 
 the bulk gauge multiplet directly communicates 
 with the SUSY breaking sector and transmits the SUSY breaking 
 to the PS brane, namely the gaugino mediation \cite{gMSB}, 
 so that the resultant soft SUSY braking mass spectrum 
 is automatically flavor blind.

\section{Setup} 
We begin with a pure SUSY SO(10) gauge theory in 5D bulk. 
The fifth dimension is compactified on the orbifold 
 $S^1/(Z_2\times Z_2^\prime) $. 
A circle $S^1$ with radius $R$ is divided by 
 a $Z_2$ orbifold transformation $y \to -y$ 
 ($y$ is the fifth dimensional coordinate $ 0 \leq y < 2 \pi R$)
 and this segment is further divided by a $Z_2^\prime$ transformation 
 $y^\prime \to -y^\prime $ with $y^\prime = y + \pi R/2$. 
There are two inequivalent orbifold fixed points at 
 $y=0$ and $y=\pi R/2$. 
Under this orbifold compactification, a general bulk 
 wave function is classified with respect to its parities,  
 $P=\pm$ and $P^\prime=\pm$, under $Z_2$ and $Z_2^\prime$, 
 respectively. 

Assigning the parity ($P,P^\prime $) as listed in Table~I, 
 only the PS gauge multiplet has zero-mode 
 and the bulk 5D N=1 SUSY SO(10) gauge symmetry is broken 
 to 4D N=1 SUSY PS gauge symmetry. 
Since all vector multiplets has wave functions  
 on the brane at $y=0$, SO(10) gauge symmetry is respected there, 
 while only the PS symmetry is on the brane at $y=\pi R/2$ 
 (PS brane).

\begin{table}[h]
\begin{center}
\begin{tabular}{|c|c|c|}
\hline
$(P,P')$ & bulk field & mass\\
\hline 
& & \\
$(+,+)$ & $V(15,1,1)$, $V(1,3,1)$, $V(1,1,3)$ & $\frac{2n}{R}$\\
& & \\
\hline
& & \\
$(+,-)$ &  $V(6,2,2)$ & $\frac{(2n+1)}{R}$ \\
& & \\
\hline
& & \\
$(-,+)$ &  $\Phi (6,2,2)$
& $\frac{(2n+1)}{R}$\\
& & \\
\hline
& & \\
$(-,-)$ & $\Phi (15,1,1)$, $\Phi (1,3,1)$, $\Phi (1,1,3)$ & $\frac{(2n+2)}{R}$ \\
& & \\
\hline
\end{tabular}
\end{center}
\caption{
 ($P,~P^\prime$) assignment and masses ($n \geq 0$) of fields 
 in the bulk SO(10) gauge multiplet $(V,~\Phi)$ 
 under the PS gauge group. 
$V$ and $\Phi$ are the vector multiplet and adjoint chiral 
 multiplet in terms of 4D N=1 SUSY theory. 
}
\label{t1}
\end{table}

In our setup, all matter and Higgs multiplets are 
 on the PS brane, where only the PS symmetry is manifest 
 so that the particle contents are in the representation 
 under the PS gauge symmetry, not necessary to be 
 in SO(10) representation.  
Thus, the particle contents do not need to 
 include harmful Higgs fields like $({\bf 6},{\bf 1},{\bf 1})$,  
 which is included in ${\bf 10}$ Higgs multiplets 
 in a SO(10) model and mediates the dimension five operator 
 relevant for proton decay,
 and there is no proton decay problem \cite{Mohapatra}. 
Even if such fields are introduced into a model in some reason, 
 they do not need to have couplings with matter multiplets. 
In fact, it is easy to impose some symmetry (parity) 
 to forbid such couplings or even if such couplings are simply 
 neglected, they are not introduce  by virtue of 
 non-renormalization theorem. 
For a different setup, see \cite{Raby}.

With respect to SU(4)$_c\times$SU(2)$_L\times$SU(2)$_R$, 
 we introduce matter multiplets, 
 the left - and right- handed quarks and leptons 
 of a given i-th generation assigned as 
\begin{eqnarray}
	\left(
		\begin{array}{cccc}
		u_r & u_y & u_b & \nu_e \\
		d_r & d_y & d_b & e
		\end{array}
	\right)_{L(R)}	\equiv F_{L(R)1} ,
\end{eqnarray}
$F_{L(R)2}$ and $F_{L(R)3}$ are likewise defined 
 for the 2nd and 3rd generations.
Their transformation properties are 
 $F_{Li}=({\bf 4},{\bf 2},{\bf 1})$ 
 and $F_{Ri}=({\bf 4},{\bf 1},{\bf 2})$, 
 so that ($F_{Li} + \overline{F_{Ri}}$ ) 
 yields the {\bf 16} of SO(10): 
 ${\bf 16}  =({\bf 4}, {\bf 2}, {\bf 1})
            +(\overline{{\bf 4}}, {\bf 1}, {\bf 2})$.

Since $({\bf 4},{\bf 2},{\bf 1})\times
(\overline{{\bf 4}},{\bf 1},{\bf 2})=
 ({\bf 1},{\bf 2},{\bf 2})+ ({\bf 15},{\bf 2},{\bf 2})$, 
 the Dirac masses for quarks and leptons can be generated by 
 $({\bf 1},{\bf 2},{\bf 2})_H$ and/or 
 $({\bf 15},{\bf 2},{\bf 2})_H$. 
We introduce the Higgs multiplets, which can works as 
 $({\bf 1},{\bf 2},{\bf 2})_H + ({\bf 15},{\bf 2},{\bf 2})_H$. 
Through the same structure as in the minimal SO(10) 
 with ${\bf 10}+\overline{\bf 126}$,  
 $({\bf 1},{\bf 2},{\bf 2})_H \subset {\bf 10}$ 
 is responsible for the $b-\tau$ unification at GUT scale, 
 while $({\bf 15},{\bf 2},{\bf 2})_H \subset \overline{\bf 126}$ 
 can ameliorate the bad relations, $m_e =m_u$ and $m_\mu =m_s$,  
 for the first two generations.

Finally, in order to break the PS symmetry to the SM one 
 and also to generate the right-handed neutrino masses, 
 we introduce Higgs multiplets of the fundamental and
 anti-fundamental representations under SU(4)$_c$. 
Here, we impose the left-right symmetry in our model, 
 namely, the model is invariant under the exchange 
 between $L \leftrightarrow R$. 
Particle contents for matter and Higgs multiplets are summarized in Table~2. 
Here we have included $({\bf 6},{\bf 1},{\bf 1})_H$ but 
 it can be decoupled to matter multiplets by imposing 
 some symmetry (parity) as we discussed in the following. 

\begin{table}[h]
{\begin{center}
\begin{tabular}{|c|c|}
\hline
& brane at $y=\pi R/2$ \\ 
\hline
\hline
& \\
Matter Multiplets & $\psi_i=F_{Li} \oplus F_{Ri} \quad (i=1,2,3)$ \\
 & \\
\hline
 & \\
Higgs Multiplets & 
$({\bf 1},{\bf 2},{\bf 2})_H$,  
$({\bf 1},{\bf 2},{\bf 2})'_H$,
$({\bf 15},{\bf 1},{\bf 1})_H$,
$({\bf 6},{\bf 1},{\bf 1})_H$ \\  & 
$({\bf 4},{\bf 1},{\bf 2})_H$, 
$(\overline{{\bf 4}},{\bf 1},{\bf 2})_H$, 
$({\bf 4},{\bf 2},{\bf 1})_H$, 
$(\overline{{\bf 4}},{\bf 2},{\bf 1})_H$  \\
& \\
\hline
\end{tabular}
\end{center}}
\caption{
Particle contents on the PS brane. 
Here, we impose the left-right symmetry. 
}
\label{localization}
\end{table}

In the following conveniences, let us introduce the following notations:
\bea
H_1&=&({\bf 1},{\bf 2},{\bf 2})_H, ~H_1^{\prime}=({\bf 1},{\bf 2},{\bf 2})'_H,
\nonumber \\
H_6&=&({\bf 6},{\bf 1},{\bf 1})_H, ~H_{15}=({\bf 15},{\bf 1},{\bf 1})_H,
\nonumber \\ 
H_L&=&({\bf 4},{\bf 2},{\bf 1})_H,
~\overline{H_L} =(\overline{{\bf 4}},{\bf 2},{\bf 1})_H,  
\nonumber \\
H_R&=&({\bf 4},{\bf 1},{\bf 2})_H,
~\overline{H_R}=(\overline{{\bf 4}},{\bf 1},{\bf 2})_H.
\eea

Superpotential relevant for fermion masses is given by%
\footnote{
For simplicity, we have introduced only minimal terms 
 necessary for reproducing observed fermion mass matrices. 
}
\bea
W_Y&=&Y_{1}^{ij}F_{Li}\overline{F_{Rj}}H_1
+\frac{Y_{15}^{ij}}{M_5} F_{Li} \overline{F_{Rj}}
 \left(H_1^{\prime} H_{15} \right) \nonumber\\ 
&+&\frac{Y_R^{ij}}{M_5}\overline{F_{Ri}}
 \overline{F_{Rj}} \left(H_R H_R \right) 
 +\frac{Y_L^{ij}}{M_5} F_{Li}F_{Lj} 
 \left(\overline{H_L} \overline{H_L} \right), 
\label{Yukawa}
\eea 
where $M_5$ is the 5D Planck scale. 
The product, $H_1^{\prime} H_{15}$, effectively works 
 as $({\bf 15},{\bf 2},{\bf 2})_H$, 
 while $H_R H_R$ and $\overline{H_L}\overline{H_L}$ 
 effectively work as $({\bf 10},{\bf 1},{\bf 3})$ and 
 $(\overline{{\bf 10}},{\bf 3},{\bf 1})$, respectively, 
 and are responsible for the left- and the right-handed 
 Majorana neutrino masses. 
Note that $Y_R$ and $Y_L$ are independent of the Dirac Yukawa
 couplings and there are a sufficient number of free parameters 
 to fit the neutrino oscillation data. 
Assuming appropriate VEVs for Higgs multiplets, 
 we can parameterize fermion mass matrix as the following form \cite{M-F-N}: 
\begin{eqnarray}
 M_u &=& c_{10} M_{1,2,2}+ c_{15} M_{15,2,2} \; , 
 \nonumber \\
 M_d &=& M_{1,2,2} + M_{15,2,2} \; ,   
 \nonumber \\
 M_D &=& c_{10} M_{1,2,2} - 3 c_{15} M_{15,2,2} \; , 
 \nonumber \\
 M_e &=& M_{1,2,2} - 3 M_{15.2,2} \; , 
 \nonumber \\
 M_L &=& c_L M_{10,3,1} \; ,
 \nonumber \\ 
 M_R &=& c_R M_{10,1,3} \; . 
\label{massmatrix}
\end{eqnarray}  
Here, the so-called Georgi-Jarlskog factor, $-3$, 
 appears in the lepton mass matrix 
 as the Clebsch-Gordan coefficient 
 associated with the basis diag$(1,1,1,-3)$ 
 for the SU(4)$_c$ adjoint Higgs 
 $({\bf 15},{\bf 2}, {\bf 2})_H$.

We introduce Higgs superpotential invariant under the PS symmetry 
 such as 
\bea
W &=& 
 \frac{m_1}{2} H_1^2 + \frac{m_1^\prime}{2} H_1^{\prime 2} 
 + m_{15}~{\rm tr}\left[H_{15}^2 \right] 
  +m_4 \left(\overline{H_L}H_L+\overline{H_R}H_R\right) \nonumber\\
&+& 
\left(H_L \overline{H_R}+ \overline{H_L} H_R \right) 
\left( \lambda_1 H_1 + \lambda_1^\prime H_1^\prime \right) 
+\lambda_{15} \left(\overline{H_R} H_R + \overline{H_L} H_L\right) 
H_{15} \nonumber\\
&+&
\lambda~{\rm tr}\left[H_{15}^3 \right]+
\lambda_6 
 \left( H_L^2+ \overline{H_L}^2 + H_R^2 + \overline{H_R}^2 \right) 
 H_6 .
\label{HiggsW}
\eea
Parameterizing 
 $ \langle H_{15} \rangle =\frac{v_{15}}{2 \sqrt{6}} 
 {\rm diag}(1,1,1,-3)$, 
 SUSY vacuum conditions from Eq.~(\ref{HiggsW}) and 
 the D-terms are satisfied by solutions,  
\bea
v_{15} =\frac{2 \sqrt{6}}{3 \lambda_{15}} m_4,~~~
\langle           H_R  \rangle = 
\langle \overline{H_R} \rangle = 
\sqrt{
 \frac{8 m_4}{3 \lambda_{15}^2} 
 \left( m_{15} -\frac{\lambda}{\lambda_{15}} m_4 \right) }
 \equiv v_{PS} 
\eea 
and others are zero, by which the PS gauge symmetry is broken 
 down to the SM gauge symmetry.  
We choose the parameters so as to be 
 $ v_{15} \simeq \langle H_R \rangle = \langle \overline{H_R} \rangle$.  
Note that the last term in Eq.~(\ref{HiggsW}) is necessary 
 to make all color triplets in $H_R$ and $\overline{H_R}$ heavy.

Weak Higgs doublet mass matrix is given by
\begin{equation}
\left(
       \begin{array}{ccc}
        H_1, & H_1^\prime, & H_L \end{array}
\right)\left(
        \begin{array}{ccc}
        m_1 &  0         & \lambda_1        \langle H_R \rangle \\
        0   & m_1^\prime & \lambda_1^\prime \langle H_R \rangle \\ 
        \lambda_1 \langle \overline{H_R} \rangle &  
        \lambda_1^\prime \langle \overline{H_R} \rangle & m_4 
        \end{array}
\right)\left(
        \begin{array}{c}
        H_1\\
        H_1^\prime \\
        \overline{H_L}
        \end{array}
\right).
\end{equation} 
In order to realize the MSSM at low energy, 
 only one pair of Higgs doublets out of the above tree pairs 
 should be light, while others have mass of the PS symmetry breaking scale. 
This doublet-doublet Higgs mass splitting requires 
 the fine tuning of parameters to satisfy 
\bea
\det M=m_1 m_1^\prime m_4 - 
 (m_1 \lambda_1^{\prime 2} + m_1^\prime \lambda_1^2) v_{PS}^2=0.
\label{vev}
\eea

\section{Gauge coupling unification}
In the orbifold GUT model, we assume that 
 the GUT model takes place at some high energy 
 beyond the compactification scale.  
For the theoretical consistency of the model, 
 the gauge coupling unification should be realized 
 at some scale after taking into account 
 the contributions of Kaluza-Klein modes 
 to the gauge coupling running \cite{Hall} \cite{GHU}. 

In our setup, the evolution of gauge coupling 
 has three stages, $G_{321}$ (SM+MSSM), $G_{422}$ (the PS) 
 and $M_c =1/R$. 
For simplicity, we assume $v_{PS}=M_c$ in our analysis. 
Furthermore, since we have imposed the left-right symmetry, 
 SU(2)$_L$ and SU(2)$_R$ gauge couplings 
 must coincide with each other at the scale $\mu=v_{PS}$. 
As a consequence, the PS scale is fixed 
 from the gauge coupling running in the MSSM stage.

\begin{figure}[h]
\begin{center}
{\includegraphics*[width=.6\linewidth]{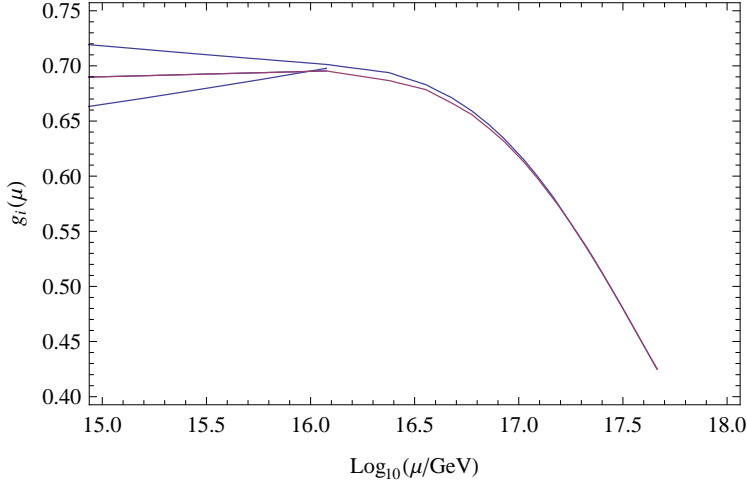}
\label{Fig}}
\caption{
Gauge coupling unification in the left-right symmetric case. 
Each line from top to bottom corresponds to 
 $g_3$, $g_2$ and $g_1$ for $ \mu < M_c$,
 while  $g_3=g_4$ and $g_2=g_{2R}$ for $ \mu > M_c$. 
}
\end{center}
\end{figure}

In the $G_{321}$ stage, we have 
\bea
\frac{1}{\alpha_i (\mu)}=\frac{1}{\alpha_i(M)}
 +\frac{1}{2\pi}b_i\mbox{ln}\left(
\frac{M}{\mu}\right); ~~(i=3,2.1), 
\eea
were $b_i$s are
\bea
b_3=-7,~b_2=-19/6,~b_1=41/10
\eea
for $M_Z <\mu < M_{SUSY}$ and
\bea
b_3=-3,~b_2=1,~b_1=33/5
\eea
for $M_{SUSY} < \mu < M_c=v_{PS}$. 
At the PS scale, the matching condition holds
\bea
\alpha_3^{-1}(M_c)&=&\alpha_4^{-1}(M_c)\nonumber\\
\alpha_2^{-1}(M_c)&=&\alpha_{2L}^{-1}(M_c)\nonumber\\
\alpha_1^{-1}(M_c)&=&[2\alpha_4^{-1}(M_c)+3\alpha_{2R}^{-1}]/5
\label{maching}
\eea
In the PS stage $\mu > M_c$, the threshold corrections 
 $\Delta_i$ due to KK mode in the bulk are added, 
\bea
\frac{1}{\alpha_i (\mu)}=\frac{1}{\alpha_i(M_c)}
+\frac{1}{2\pi}b_i\mbox{ln}\left(
\frac{M_c}{\mu}\right)+\Delta_i.~~(i=4,2_L,2_R)
\eea
The beta functions from the matter and Higgs multiplets 
 on the PS brane are 
\be
 b_4=3, ~b_{2L}=b_{2R}=6 .
\label{lrs}
\ee
KK mode contributions are given by 
\bea
\Delta_i&=& \frac{1}{2\pi}b_i^{even}\sum_{n=0}^{N_l}
\theta(\mu-(2n+2)M_c)\mbox{ln}\frac{(2n+2)M_c}{\mu} \nonumber\\  
&+&
\frac{1}{2\pi}b_i^{odd}\sum_{n=0}^{N_l}
\theta(\mu-(2n+1)M_c)\mbox{ln}\frac{(2n+1)M_c}{\mu}
\eea
with 
\bea
b_i^{even}&=&(-8,-4,-4) , \nonumber\\ 
b_i^{odd}&=&(-8,-12,-12)  
\eea
under $G_{422}$. 

Fig.~1 shows the gauge coupling unification 
 for the left-right symmetric case. 
The PS (compactification) scale, $M_c$, is determined 
 from the gauge coupling running 
 in the MSSM stage by imposing the matching condition, 
 $\alpha_2^{-1}(M_c)= \alpha_{2R}^{-1}(M_c) 
 =(5\alpha_1^{-1}(M_c)-2\alpha_3^{-1}(M_c))/3$, 
 and we find 
\bea 
 v_{PS}=M_c = 1.2\times 10^{16} \mbox{GeV} . 
\eea 
 for the inputs, 
 $(\alpha_1(M_Z), \alpha_2(M_Z), \alpha_3(M_Z))
 = (0.01695, 0.03382, 0.1176)$ and $M_{SUSY}=1$ TeV. 
For the scale $\mu > M_c$, there are only two independent 
 gauge couplings $\alpha_4$ and $\alpha_2=\alpha_{2R}$, 
 and so the gauge coupling unification is easily realized. 
We find the unification scale as 
\bea 
  M_{GUT} = 4.6 \times 10^{17} \mbox{GeV} . 
\eea  
As mentioned before, we assume that a more fundamental 
 SO(10) GUT theory takes place at $M_{GUT}$, 
 and it would be natural to assume $M_{GUT} \sim M_5$. 
In fact, the relation between 4D and 5D Planck scales, 
 $M_5^3/M_c \simeq M_P^2$ 
 ($M_P=2.4 \times 10^{18}$ GeV is the reduced Planck scale), 
 supports this assumption 
 with $M_c=1.2 \times 10^{16}$ GeV. 
When we abandon the left-right symmetry, 
 there is more freedom for the gauge coupling unification 
 with two independent parameters $v_{PS}$ and $M_c$.

\section{Supersymmetry breaking mediation} 
The origin of SUSY breaking and its mediation to the MSSM sector 
 is still an open question of SUSY models and 
 there have been many scenarios proposed. 
A mechanism which naturally transmits SUSY breaking 
 in a flavor-blind way is the most favorable one. 
Here we consider such a scenario. 

In higher dimensional models, the sequestering \cite{RS-SUSY} 
 is the easiest way to suppress flavor dependent SUSY breaking 
 effects to the matter sector. 
Since all matters reside on the PS brane in our model, 
 the sequestering scenario is automatically realized 
 when we simply assume a SUSY breaking sector 
 on the brane at $y=0$. 
The SO(10) gauge multiplet is in the bulk and 
 can directly communicate with the SUSY breaking sector. 
Here, we first consider the higher dimensional operator of the form,  
\bea 
 {\cal L} =\delta (y) 
 \int d^2 \theta \lambda \; 
  \frac{X}{M_5^2} 
 {\rm tr} \left[ {\cal W}^\alpha {\cal W}_\alpha \right],   
\eea 
where $\lambda$ is a dimension-less constant, and 
 $X$ is a singlet chiral superfield 
 which breaks SUSY by its F-component VEV, $X= \theta^2 F_X$. 
Therefore, the bulk gaugino first obtains SUSY breaking masses, 
\bea 
 M_\lambda = \frac{\lambda F_X M_c}{M_5^2}
 \simeq \frac{\lambda F_X M_5}{M_P^2} ,  
\eea
where $M_c$ comes from the wave function normalization 
 of the bulk gaugino, 
 and we have used the relation $M_5^3/M_c \simeq M_P^2$ 
 in the last equality. 
As usual, we take $M_\lambda =$100 GeV-1 TeV.  
With this non-zero gaugino mass at high scale, 
 SUSY breaking mass terms of sfermions are automatically 
 generated through the renormalization group equation (RGE) 
 from the compactification scale to the electroweak scale. 
Importantly, the sfermion masses generated in this way 
 are flavor blind, because the interaction transmitting 
 the gaugino mass to sfermion masses is the gauge interaction. 
This scenario is nothing but the gaugino mediation \cite{gMSB}. 
Comparing the gaugino mass to gravitino mass 
 $m_{3/2} \simeq F_X/M_P$, a typical gaugino mass 
 is smaller than the gravitino mass 
 by a factor $\lambda M_5/M_P \sim 0.1 \lambda$.

If there are some extra bulk multiplets coupling with 
 both the SUSY breaking and the MSSM sectors, 
 flavor-dependent sfermion masses can be, in general, induced. 
Thus, it is important to check whether such flavor-dependent terms 
 are small enough compared to the gaugino mediation contribution. 
Introducing an extra bulk hypermultiplet ($H_0$), let us consider 
 effective Kahler potentials both on the PS brane and the other brane, 
\bea 
 {\cal L}=
  \delta(y) \int d^4 \theta \frac{H_0^\dagger H_0  X^\dagger X}{M_5^3} 
+ \delta(y-\pi R/2) 
  \int d^4 \theta c_{ij} \frac{H_0^\dagger H_0  Q_i^\dagger Q_j}{M_5^3},  
\eea 
where $Q_i$ stands for a MSSM matter multiplet with the generation index $i$, 
 and $c_{ij}$ is a flavor-dependent dimensionless coefficient. 
Thus, one-loop corrections through $H_0$ lead to flavor-dependent 
 contributions to sfermion masses, which are roughly estimated as 
\bea 
 \Delta \tilde{m}^2_{ij} \sim 
 \frac{c_{ij}}{16 \pi^2} \frac{F_X^2 M_c^4}{M_5^6}.  
\eea 
Comparing this to the flavor-blind sfermion mass squareds 
 induced by the gaugino mass, $\tilde{m}^2 \sim M_{\lambda}^2$, 
 we find 
\bea
 \frac{\Delta \tilde{m}^2_{ij}}{\tilde{m}^2} 
 \sim \frac{c_{ij}}{16 \pi^2} \left( \frac{M_c}{M_5} \right)^2 
 \sim \frac{c_{ij}}{16 \pi^2} \left( \frac{M_c}{M_{GUT}} \right)^2 
 \simeq 4.3 \times 10^{-6} c_{ij},   
\eea
which is negligibly small.

In the simple setup, it turns out that stau 
 is the lightest superpartner (LSP), 
 which is problematic in cosmology. 
It has been found that when $M_c > M_{GUT}$, 
 the RGE running in a unified theory pushes up 
 stau mass and leads neutralino to be the LSP \cite{gMSB2}. 
However, in our model, we cannot take such an arrangement, 
 because $M_c$ and $M_{GUT}$ are fixed as $M_c < M_{GUT}$ 
 to realize the gauge coupling unification. 
In order to avoid this problem, 
 we need to extend the SUSY breaking sector. 
It is possible to introduce the gauge mediation \cite{GMSB} 
 on the PS brane, in which gravitino is normally the LSP. 
In general, we can introduce the messenger sector 
 on the brane at $y=0$. 
This setup is basically the same as in Ref.~\cite{MP}, 
 where the gauge mediation was calculated in 5D 
 with the messenger sector on one brane, 
 sfermions on the other brane 
 and gauge multiplets in the bulk. 
When the messenger scale is larger than the compactification scale
 ($M_{mess} > M_c$), 
 the gaugino mass is given by the same formula as in 4D, 
\bea 
 M_\lambda \simeq \frac{\alpha_{GUT}}{4 \pi} \frac{F_X}{M_{mess}}, 
\eea 
while sfermion masses are roughly given by 
\bea 
 \tilde{m}^2 \simeq M_\lambda^2 \left(\frac{M_c}{M_{mess}} \right)^2.  
\eea 
The sfermion mass squared is suppressed relative to 
 the gaugino mass by a geometric factor $M_c/M_{mess}$, 
 at the messenger scale. 
At low energy, sfermion masses comparable 
 to the gaugino mass are generated through the RGE running. 
In this setup, we find 
\bea 
 \frac{m_{3/2}}{M_\lambda} \simeq 
 \frac{M_{mess}}{\frac{\alpha_{GUT}}{4 \pi}M_P} 
 \gtrsim 10 
\eea 
for $M_{mess} \geq M_c$. 
Thus, in oder to have gravitino the LSP, 
 the messenger scale should be smaller 
 than the compactification scale, namely, 
 $M_{mess} \lesssim 10^{15}$ GeV.  
In this case, soft mass formulas are reduced into 
 the usual four dimensional ones 
 in the gauge mediation scenario.

\section{Conclusion}
We have proposed a simple SO(10) model 
 in five dimensions with the 5th dimension compactified on 
 the orbifold $S^1/(Z_2\times Z_2^\prime)$. 
Due to the orbifold boundary conditions, 
 a bulk N=2 SUSY and SO(10) gauge symmetry 
 are broken down to 4D N=1 SUSY PS model 
 with the gauge group, SU(4)$_c\times$SU(2)$_L\times$SU(2)$_R$.
All matter and Higgs multiplets reside only on the PS brane, 
 while the gauge multiplet is in 5D bulk.
The PS symmetry is broken to the SM one  
 by the usual Higgs mechanism on the PS brane. 
Imposing the left-right symmetry, 
 the gauge coupling unification is realized 
 at $M_{\rm GUT} \simeq 4.6 \times 10^{17}$ GeV 
 with the compactification $M_c \simeq 1.2 \times 10^{16}$ GeV. 
There are various possibilities for SUSY breaking. 
When we assume SUSY breaking on the brane at $y=0$, 
 SUSY breaking is transmitted into the PS brane 
 through the gaugino mediation.

\section*{Acknowledgments}
We would like to thank Naoyuki Haba and Nobuhito Maru 
 for valuable discussions. 
The work of N.O. is supported in part by 
 the Grant-in-Aid for Scientific Research from the Ministry 
 of Education, Science and Culture of Japan (\#18740170). 

\newpage

\end{document}